\documentclass[twocolumn,amsmath,amssymb,aps,superscriptaddress]{revtex4-2}
\usepackage{package}

\begin{document}
\title{Zero-level $CCZ$ Distillation}
\author{Tomohiro Itogawa}
\email{itogawa.tomohir@fujitsu.com}
\affiliation{%
  Graduate School of Engineering Science, Osaka University, 1-3 Machikaneyama, Toyonaka, Osaka 560-8531, Japan
}
\affiliation{%
Graduate School of Informatics, Kyoto University, Sakyo-ku, Kyoto, 606-8501, Japan
}
\affiliation{%
  Quantum Laboratory, Fujitsu Research, Fujitsu Limited, 4-1-1 Kawasaki, Kanagawa 211-8588, Japan
}

\author{Yutaka Hirano}
\email{yutaka.hirano@nano-qt.com}
\affiliation{%
Graduate School of Engineering Science, Osaka University, 1-3 Machikaneyama, Toyonaka, Osaka 560-8531, Japan
}%
\affiliation{%
Nanofiber Quantum Technologies, Inc. (NanoQT), 1-22-3 Nishiwaseda, Shinjuku-ku, Tokyo 169-0051, Japan
}%

\author{Yutaro Akahoshi}
\email{akahoshi.yutaro@fujitsu.com}
\affiliation{%
  Quantum Laboratory, Fujitsu Research, Fujitsu Limited, 4-1-1 Kawasaki, Kanagawa 211-8588, Japan
}
\affiliation{%
Fujitsu Quantum Computing Joint Research Division, Center for Quantum Information and Quantum Biology, Osaka University, 1-2 Machikaneyama, Toyonaka, Osaka, 565-8531, Japan
}
\author{Keisuke Fujii}%
\email{fujii@qc.ee.es.osaka-u.ac.jp}
\affiliation{%
  Graduate School of Engineering Science, Osaka University, 1-3 Machikaneyama, Toyonaka, Osaka 560-8531, Japan
}%
\affiliation{%
Graduate School of Informatics, Kyoto University, Sakyo-ku, Kyoto, 606-8501, Japan
}
\affiliation{%
Fujitsu Quantum Computing Joint Research Division, Center for Quantum Information and Quantum Biology, Osaka University, 1-2 Machikaneyama, Toyonaka, Osaka, 565-8531, Japan
}
\affiliation{%
  RIKEN Center for Quantum Computing (RQC), Hirosawa 2-1, Wako, Saitama 351-0198, Japan
}%

\date{\today} 

\begin{abstract}
Magic state distillation is a key component of fault-tolerant quantum computation, as it enables the implementation of non-Clifford gates such as the $T$ gate and the $CCZ$ gate via gate teleportation.
However, conventional distillation protocols require a large number of logical qubits and introduce substantial spatial and temporal overhead, posing a significant bottleneck for scalable fault-tolerant quantum computation.
In this work, we propose a zero-level distillation protocol that efficiently generates a high-fidelity logical $CCZ$ magic state using only physical qubits on a two-dimensional square lattice with nearest-neighbor interactions.
Our method leverages the transversal $T/T^\dagger$ operation of the $\llbracket 8,3,2 \rrbracket$ code to fault-tolerantly encode the state $\overline{CCZ}\ket{+++}$, which is subsequently teleported to three surface-code logical qubits via lattice surgery.
To enable teleportation between codes with different distances, we introduce adaptively initialized teleportation (AIT), a tailored initialization procedure for the surface code.
Numerical simulations demonstrate that the logical error rate scales as $p_L \simeq 300 \times p^2$ with respect to the physical error rate $p$.
For example, the proposed method improves the logical error rate by approximately one and two orders of magnitude at $p = 10^{-3}$ and $p = 10^{-4}$, respectively, compared to conventional seven-$T$-gate approaches.
The distillation circuit requires only 22 physical qubits, 3 logical qubits, and a circuit depth of 24, reducing the space-time overhead by a factor of approximately 5--10 compared to previous methods.
This result highlights the practicality of $CCZ$-state distillation in early fault-tolerant quantum computation and offers a new direction toward resource-efficient physical-level magic state distillation beyond conventional $T$-state generation.
\end{abstract}
\maketitle

\section{INTRODUCTION}
Quantum computers are expected to offer computational advantages for certain problems that are intractable on classical computers.
Prominent examples include Shor’s factoring algorithm~\cite{shor1999polynomial}, Grover’s search algorithm~\cite{grover1996fast}, and the Harrow-Hassidim-Lloyd (HHL) algorithm~\cite{harrow2009quantum}.
However, physical qubits are highly susceptible to errors arising from interactions with their environment, and thus the realization of large-scale quantum algorithms requires fault-tolerant quantum computation~\cite{shor1995scheme}.
Accordingly, the development of fault-tolerant quantum computation (FTQC) based on quantum error-correcting codes is essential and remains a central challenge in achieving practical quantum computing~\cite{fujii2015quantum}.

The surface code~\cite{kitaev1997quantum,kitaev2003fault} is one of the most promising quantum error-correcting codes.
Its compatibility with nearest-neighbor interactions on a two-dimensional lattice makes it particularly attractive for physical implementation such as superconducting qubits.
It also exhibits a high error threshold~\cite{fowler2012surface,raussendorf2001one}, and recent experiments have demonstrated operation below this threshold~\cite{google2025quantum}.
While Clifford gates can be implemented efficiently within the surface code, non-Clifford gates such as the $T$ and $CCZ$ gates remain challenging to realize~\cite{eastin2009restrictions}.
These gates are typically implemented via magic state distillation (MSD)~\cite{bravyi2005universal}, which produces high-fidelity magic states from multiple noisy ones using only Clifford operations.
However, MSD requires a large number of logical qubits, leading to substantial space-time overhead and constituting a major bottleneck for fault-tolerant quantum computation~\cite{gidney2021factor}.

To tackle this issue, physical-level distillation~\cite{goto2016minimizing,chamberland2019fault,chamberland2020very,itogawa2025efficient,hirano2024leveraging,gidney2024magic,hirano2025efficient,honciuc2024implementing} has recently attracted considerable interest.
Unlike conventional MSD performed at the logical level, this approach designs distillation circuits at the physical level by explicitly accounting for errors arising from Clifford operations, enabling the generation of high-fidelity magic states with reduced space-time overhead.
In particular, ``zero-level distillation''~\cite{itogawa2025efficient} outputs high fidelity magic states encoded in the surface code under realistic hardware constraints with a nearest-neighbor qubit array at physical error rates on the order of $10^{-3}$.
Building on this approach, several extensions have been proposed, including magic state cultivation~\cite{gidney2024magic} and related protocols~\cite{chen2025efficient, vaknin2025efficient, sahay2025fold, hirano2025efficient}.

In this work, we propose a zero-level distillation protocol for preparing $CCZ$ magic states.
We consider hardware constraints motivated by superconducting qubit architectures, where qubits are arranged on a two-dimensional square lattice.
Our zero-level $CCZ$ distillation protocol employs the logical $CCZ$ operation implemented via transversal $T/T^\dagger$ gates of the $\llbracket 8,3,2 \rrbracket$ code, enabling a fault-tolerant encoding of the state $\overline{CCZ}\ket{+++}$ ~\cite{kubica2015unfolding, The_smallest_interesting_colour_code, chen2019machine}.
Once this encoding is successfully completed, we employ lattice surgery to teleport the encoded logical state into three surface-code logical qubits, thereby producing a high-fidelity $CCZ$ magic state.
To enable teleportation between codes with different distances, we introduce adaptively initialized teleportation (AIT), a tailored initialization procedure for the surface code.
Specifically, because the $\llbracket 8,3,2 \rrbracket$ code possesses only weight-2 logical $Z$ operators, it is not directly compatible with the surface code, which typically has long logical $Z$ operators.
We therefore initialize the surface code so that it temporarily supports short logical $Z$ operators, enabling lattice surgery between the two codes and facilitating the teleportation step.

Numerical simulations demonstrate that the logical error rate scales quadratically with the physical error rate, approximately following $p_L \simeq 300 \times p^2$.
Compared with the conventional implementation of a $CCZ$ gate using seven $T$ gates, the proposed protocol achieves an improvement in logical error rate by roughly one order of magnitude at $p = 10^{-3}$ and two orders of magnitude at $p = 10^{-4}$.
The distillation protocol also attains a high success probability, reaching approximately 30\% at $p = 10^{-3}$ and 90\% at $p = 10^{-4}$.
The distillation circuit consists of 22 physical qubits and three logical qubits, and has a very shallow circuit depth of 24.
As a result, the space-time overhead is reduced by a factor of approximately 5–10 compared to conventional approaches.

These results demonstrate the practicality of $CCZ$-state distillation for early fault-tolerant quantum computation (early-FTQC) and provide a new direction toward resource-efficient, physical-level magic state distillation beyond traditional $T$-state-based schemes.
Moreover, while prior research has primarily focused on distillation of $T$ states, direct $CCZ$-state distillation remains comparatively underdeveloped; our approach offers a complementary route for generating non-Clifford resources.

The remainder of this paper is organized as follows.
Section \ref{sec:preliminary} provides an overview of the $\llbracket 8,3,2 \rrbracket$ code.
Section \ref{sec:zero_level_ccz_distillation} presents the proposed zero-level $CCZ$ distillation protocol in detail.
Section \ref{sec:numerical_simulation} reports the results of our numerical simulations.
Finally, Section \ref{sec:conclusion} concludes the paper.

\section{PRELIMINARY}
  \label{sec:preliminary}
\begin{figure}[t]
    \centering
    \includegraphics[keepaspectratio, scale=0.45]{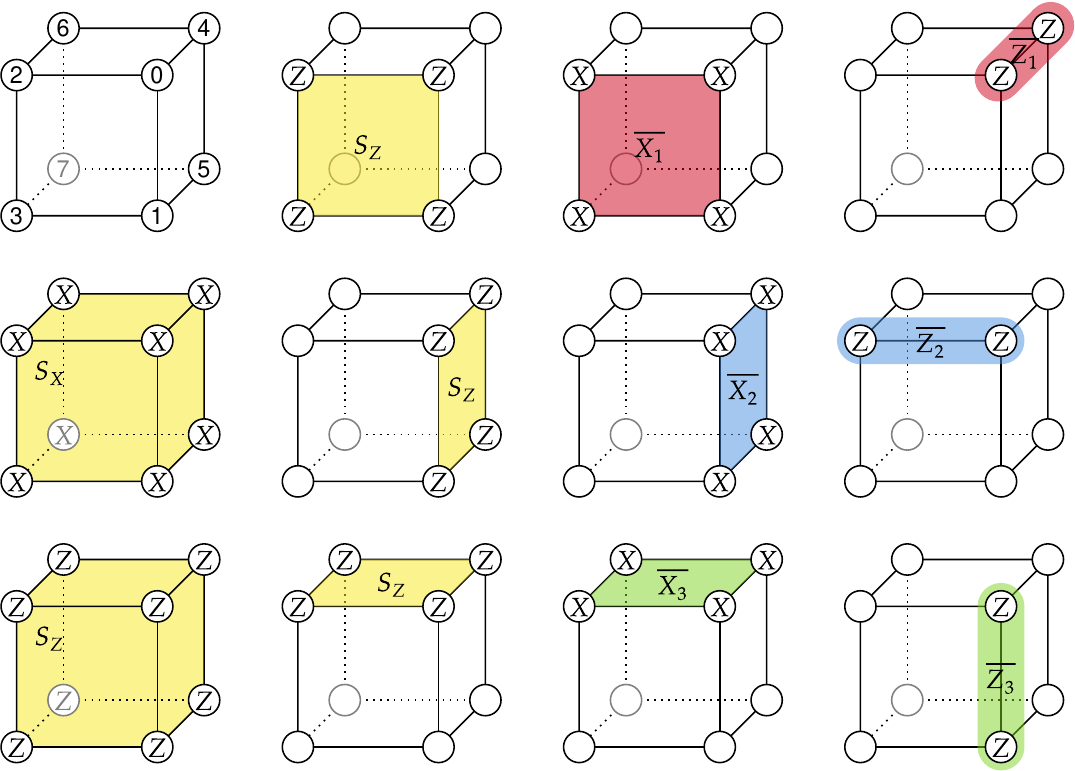}
    \caption{
    \justifying
    Geometric representation of the $\llbracket 8,3,2 \rrbracket$ code.
    Physical qubits are located at the vertices of the cube.
    The yellow cells and faces represent the $X$-type and $Z$-type stabilizer generators, respectively.
    The red, blue, and green faces and edges correspond to the logical operators.
    }
    \label{fig:8_3_2_code}
  \end{figure}

  In this section, we provide a brief overview of the $\llbracket 8,3,2 \rrbracket$ code~\cite{kubica2015unfolding, The_smallest_interesting_colour_code, chen2019machine}.
  The $\llbracket 8,3,2 \rrbracket$ code is a CSS code that encodes three logical qubits into eight physical qubits and has a code distance of two.
  Its stabilizer generators are given, as illustrated in Fig.~\ref{fig:8_3_2_code}, by
  \begin{gather*}
  S_{X,1} = X^{\otimes 8}, ~ S_{Z,1} = Z^{\otimes 8},\\
  S_{Z,2} = Z_0 Z_1 Z_2 Z_3, ~ S_{Z,3} = Z_0 Z_1 Z_4 Z_5, ~ S_{Z,4} = Z_0 Z_2 Z_4 Z_6,
  \end{gather*}
  where $X_i$ and $Z_i$ denote the Pauli $X$ and $Z$ operators acting on the $i$-th physical qubit, respectively.
  The logical operators are defined as
  \begin{gather*}
  L_{X,1} = X_0 X_1 X_2 X_3, ~ L_{X,2} = X_0 X_1 X_4 X_5, ~ L_{X,3} = X_0 X_2 X_4 X_6,\\
  L_{Z,1} = Z_0 Z_4, ~ L_{Z,2} = Z_0 Z_2, ~ L_{Z,3} = Z_0 Z_1.
  \end{gather*}
  
  It is known that the $\llbracket 8,3,2 \rrbracket$ code is one of the smallest triorthogonal codes that support transversal entangling non-Clifford gates~\cite{The_smallest_interesting_colour_code, chen2019machine}.
  In particular, this code admits a transversal implementation of the $CCZ$ gate by applying $T$ and $T^\dagger$ gates to each physical qubit in a specific pattern.
  Explicitly,
  \begin{align}
    \overline{CCZ} = T_0 T_1^\dagger T_2^\dagger T_3 T_4^\dagger T_5 T_6 T_7^\dagger.
  \end{align}
  
\section{ZERO-LEVEL $CCZ$ DISTILLATION}
\label{sec:zero_level_ccz_distillation}

\begin{figure*}[tbp]
\centering
\includegraphics[keepaspectratio, width=\linewidth]{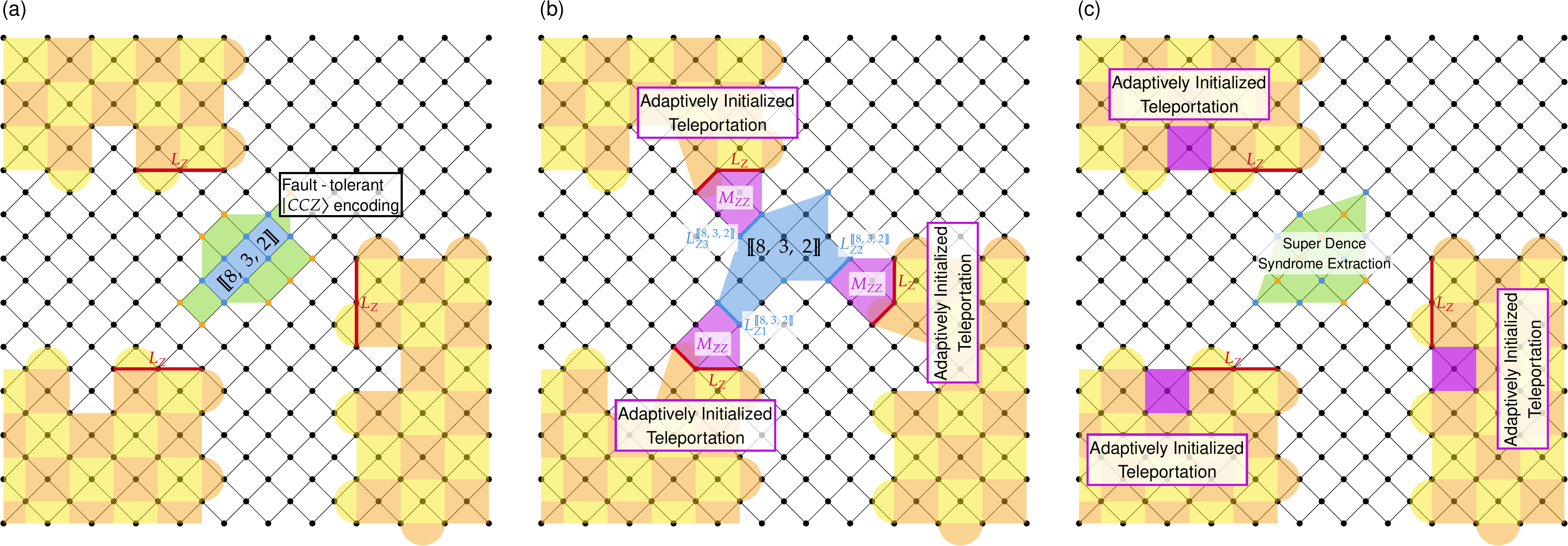}
\caption{
\justifying
  Overview of the proposed zero-level $CCZ$ state distillation protocol.
  (a) Using a non-fault-tolerant encoding circuit and transversal CNOT gates between ancilla blocks, a $|CCZ\rangle$ state is prepared in the $\llbracket 8,3,2 \rrbracket$ code. Additionally, intermediate surface-code states are prepared for adaptively initialized teleportation (AIT).
  (b) AIT is performed by executing lattice surgery between the two-body logical operators of the $\llbracket 8,3,2 \rrbracket$ code and the three-body logical operators of the intermediate state.
  (c) While performing superdense syndrome extraction on the $\llbracket 8,3,2 \rrbracket$ code, the surface codes are completed through stabilizer measurements, thereby finalizing AIT.
}
\label{fig:overview}
\end{figure*}

\begin{figure}[tbp]
  \centering
  \includegraphics[scale = 0.5]{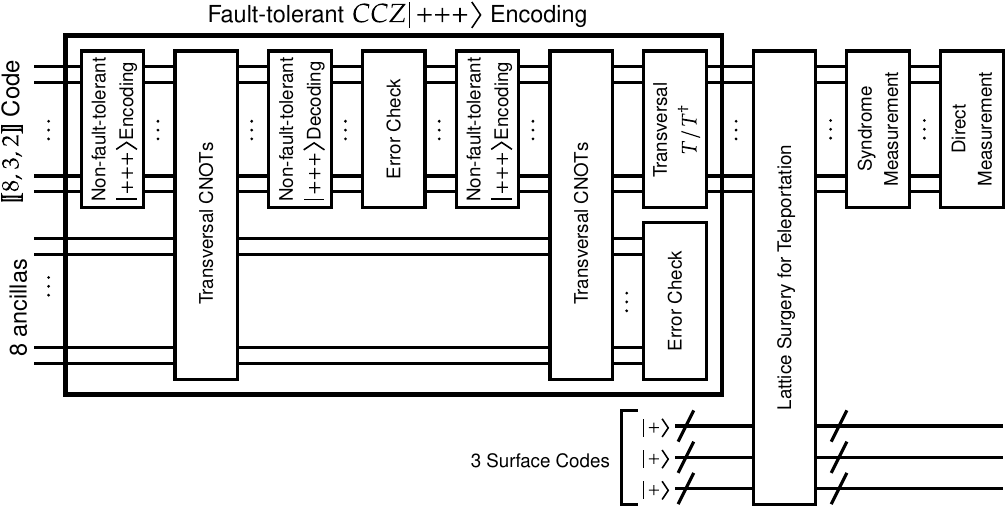}
  \caption{
  \justifying
    Overall structure of zero-level $CCZ$ state distillation.
    A non-fault-tolerant encoding/decoding circuit is used to prepare the state $\ket{+++}$, and a verification procedure is then applied to obtain a fault-tolerant encoding.
    Transversal $T/T^\dagger$ gates are subsequently applied to implement the $CCZ$ operation, after which the three logical qubits are teleported to three surface-code logical qubits.}
  \label{fig:ccz_flow}
\end{figure}

Zero-level $CCZ$ state distillation is a distillation method with very low space-time overhead, achieved by performing the distillation of the $\overline{CCZ}\ket{+++}$ state entirely at the physical-qubit level.
An overview of the proposed scheme is shown in Fig.~\ref{fig:overview}, and a step-by-step view is provided in Fig.~\ref{fig:ccz_flow}.
The protocol consists of two main stages: the state $\overline{CCZ}\ket{+++}$ is first encoded using the $\llbracket 8,3,2 \rrbracket$ code fault-tolerantly, and the resulting encoded state is then teleported into three surface-code logical qubits.
The explicit circuit construction is provided in Appendix \ref{sec:circuit}.
The resulting construction uses only 22 physical qubits for the distillation circuit and three logical surface-code qubits for the output, with a total circuit depth of 24. This depth corresponds to three rounds of surface-code syndrome extraction, highlighting the low temporal overhead of the proposed protocol.
Because zero-level distillation is carried out at the physical layer rather than the logical layer, deeper circuits lead to a rapid increase in the number of fault paths and therefore cause a significant deterioration in performance.
Achieving a shallow circuit depth is thus crucial.
In addition, the distillation circuit must operate using only nearest-neighbor connectivities on a two-dimensional square lattice, and the circuit must be designed to satisfy this architectural constraint.

\subsection{Fault-tolerant $\overline{CCZ}\ket{+++}$ encoding circuit}
\begin{figure}[tbp]
    \centering
    \includegraphics[keepaspectratio, scale=0.7]{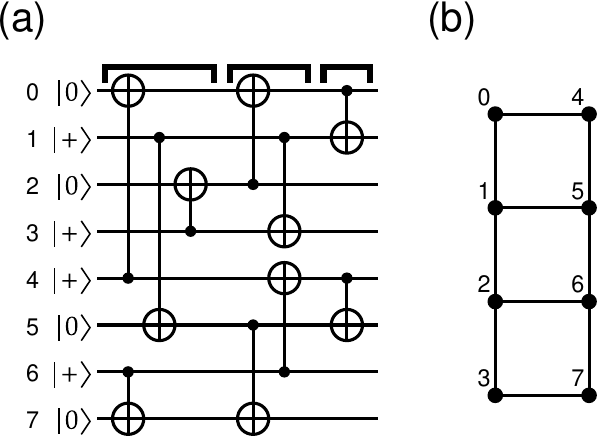}
    \caption{
    \justifying
    (a) Non-fault-tolerant $\ket{+++}$ encoding circuit, consisting of a depth-3 minimal sequence of CNOT gates.
    (b) Required qubit connectivity for the encoding circuit, compatible with strictly nearest-neighbor interactions on a two-dimensional square lattice.}
    \label{fig:non_ft_encoding}
  \end{figure}

The fault-tolerant encoding circuit for the state $\overline{CCZ}\ket{+++}$ is constructed on top of a non-fault-tolerant encoding circuit for $\ket{+++}$.
The non-fault-tolerant circuit is shown in Fig.~\ref{fig:non_ft_encoding}(a).
This circuit is a minimal construction consisting of eight CNOT gates with a circuit depth of 3, and it encodes the state $\ket{+++}$ into the $\llbracket 8, 3, 2 \rrbracket$ code.
It takes four $\ket{0}$ states and four $\ket{+}$ states as inputs, which correspond to the four $Z$-stabilizers, one $X$-stabilizer, and three logical $X$-operators of the code.
Moreover, as shown in Fig.~\ref{fig:non_ft_encoding}(b), all required qubit interactions can be implemented using nearest-neighbor connectivities on a two-dimensional square lattice.

\begin{figure*}[tbp]
    \centering
    \includegraphics[keepaspectratio, scale=0.55]{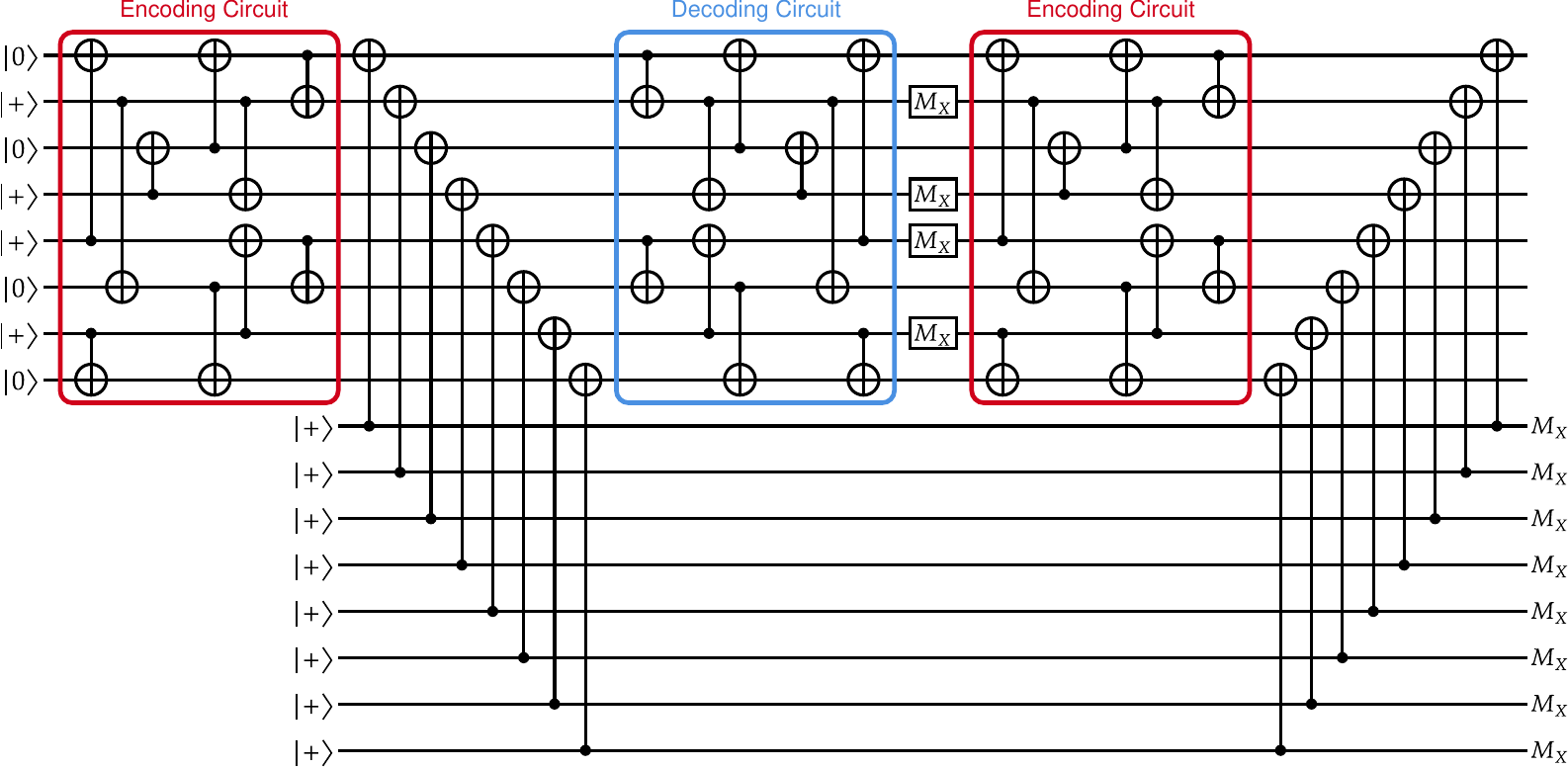}
    \caption{
    \justifying
    Fault-tolerant $\ket{+++}$ encoding circuit.
    The state is first encoded using the non-fault-tolerant circuit, followed by a transversal CNOT applied to ancilla qubits.
    The circuit is then decoded, and the four data qubits that were initially prepared in the $\ket{+}$ state are measured in the $X$-basis for verification.
    A second round of encoding and transversal CNOT is performed, and the output is accepted only when all measurement outcomes are $+1$.}
    \label{fig:ft_encoding}
  \end{figure*}

For the fault-tolerant encoding of the $\ket{+++}$ state in the $\llbracket 8,3,2 \rrbracket$ code, several approaches have been proposed for fault-tolerant logical-state preparation, including stabilizer-measurement-based methods and flag-qubit-assisted constructions~\cite{honciuc2024implementing}.
However, under nearest-neighbor constraints on a two-dimensional square lattice, these approaches are difficult to implement since they require many ancilla qubits and deep circuits.
We therefore propose a new state-preparation method based on transversal CNOT operations and decoding.
This method is inspired by the double-check circuit used in magic state cultivation~\cite{gidney2024magic}.
Fig.~\ref{fig:ft_encoding} illustrates the procedure for obtaining a fault-tolerant encoding of $\ket{+++}$ using the non-fault-tolerant encoding circuit.
This method is well suited to stringent constraints such as those considered here, because the ancilla block can be prepared as the product state $\ket{+}^{\otimes 8}$ and only simple connectivity is required.

We begin by encoding $\ket{+++}$ using the non-fault-tolerant circuit.
A transversal CNOT is then applied from the encoded block to eight ancilla qubits initialized in the state $\ket{+}$.
Because the $\llbracket 8,3,2 \rrbracket$ code is a CSS code, this operation leaves the $X$-type stabilizers and the logical $X$ operators invariant.
Next, we execute the encoding circuit in reverse order to decode the state, and we perform $X$-basis measurements on the four qubits that were initially prepared in $\ket{+}$.
These measurement outcomes correspond to the $X$-type stabilizer generator and the three logical $X$-operators; in the absence of errors, all outcomes should be $+1$.
If any outcome yields $-1$, an error is detected and the output is discarded.
The same sequence, encoding followed by a transversal CNOT, is repeated once more.
This renders the circuit symmetric, and ensures that the overall transformation is the identity when no faults occur.
Consequently, the ancilla qubits are guaranteed to return to the state $\ket{+}$ in the error-free case, and fault tolerance is obtained by discarding the output only when a $-1$ measurement is observed.
This circuit provides fault tolerance against $Z$-errors, though not against $X$-errors.
However, for the preparation of $\ket{+++}$, logical $X$ errors are harmless, and other $X$-type errors are detected and removed by postselection based on subsequent syndrome measurements.
Once a fault-tolerant preparation of $\ket{+++}$ is obtained, we apply the transversal $T/T^\dagger$ gates to produce the desired state $\overline{CCZ}\ket{+++}$.

\subsection{Teleportation into the Surface Codes}
\begin{figure}
  \centering
  \includegraphics[keepaspectratio, scale=1]{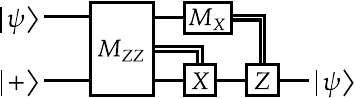}
  \caption{
  \justifying
    Basic circuit for teleportation using simultaneous measurements.
    A $ZZ$ measurement is performed between the data qubit in the state $\ket{\psi}$ and an ancilla qubit prepared in $\ket{+}$, followed by an $X$-basis measurement of the data qubit, transferring the quantum state to the ancillary qubit.
    In lattice-surgery-based teleportation, this $ZZ$ measurement can be performed between qubits belonging to different codes.
    }
  \label{fig:teleport_basic}
\end{figure}

Although the $\llbracket 8,3,2 \rrbracket$ code is well suited for implementing the $CCZ$ gate, it is difficult to handle on a two-dimensional square lattice and provides limited fault tolerance due to its distance of~2.
To overcome these limitations, the three encoded logical qubits should be teleported into surface-code logical qubits, which are easier to implement on a planar architecture and can achieve substantially larger code distances.
The basic teleportation circuit is shown in Fig.~\ref{fig:teleport_basic}~\cite{poulsen2017fault}.
In this circuit, a $ZZ$ measurement is performed between a data qubit in the state $\ket{\psi}$ and an ancilla qubit prepared in $\ket{+}$, followed by an $X$-basis measurement of the data qubit.
This sequence teleports the quantum state to the ancilla qubit.
In lattice-surgery-based teleportation, this $ZZ$ measurement can be performed between qubits belonging to different codes.

\begin{figure}
  \centering
  \includegraphics[keepaspectratio, scale=0.7]{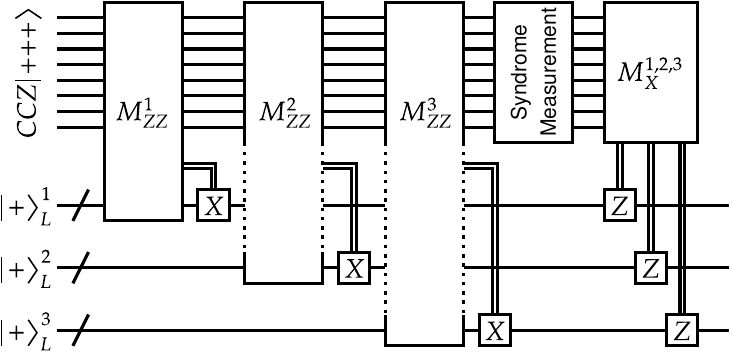}
  \caption{
  \justifying
    Teleportation circuit from the $\llbracket 8,3,2 \rrbracket$ code to three surface-code logical qubits.
    A logical $ZZ$ measurement is performed independently for each logical qubit, and after syndrome extraction, the $\llbracket 8,3,2 \rrbracket$ block is directly measured in the $X$-basis.}
  \label{fig:teleport_ccz}
\end{figure}

In our protocol, this circuit is applied independently to each of the three logical qubits to teleport them from the $\llbracket 8,3,2 \rrbracket$ code to the three surface codes.
After performing the three logical $ZZ$ measurements, the qubits in the $\llbracket 8,3,2 \rrbracket$ block are directly measured in the $X$ basis.
The complete procedure is illustrated in Fig.~\ref{fig:teleport_ccz}.

\begin{figure*}[htbp]
  \centering
  \includegraphics[keepaspectratio, scale=0.35]{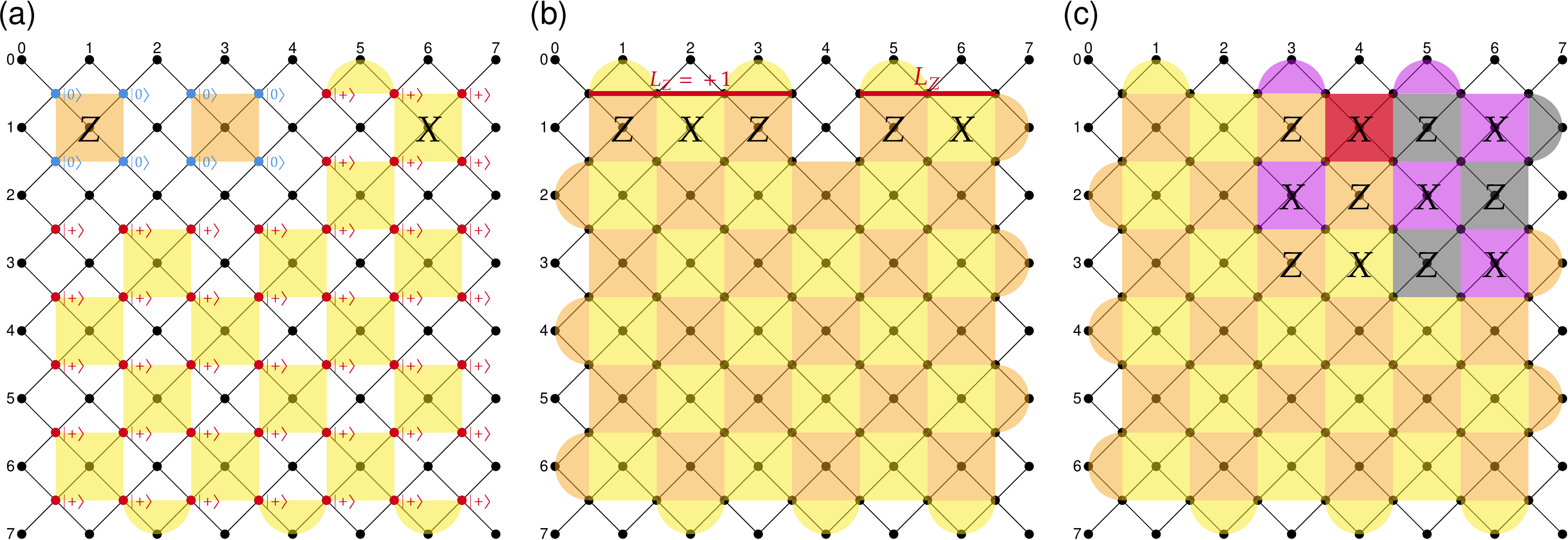}
  \caption{
  \justifying
  Initialization procedure of a surface-code $\ket{+}$ state for lattice surgery with a small-distance code.
  (a) Prepare a product state in which part of the qubits are initialized to $\ket{0}$ and the rest to $\ket{+}$.
  (b) Measure stabilizers to obtain an intermediate state suitable for a logical $ZZ$ measurement. 
  In this step, one stabilizer is left unmeasured so that the logical $Z$ operator becomes a three-body operator.
  The red lines indicate a distance-3 logical $Z$ operator inherited from the $\ket{+}$ region and a logical $Z$ operator with a fixed measurement outcome inherited from the $\ket{0}$ region.
    (c) Measure the previously unmeasured stabilizer (red face) to complete the surface code. The gray $Z$ stabilizers are those associated with the distance-3 surface-code patch in the upper-right region.
    The purple $X$ stabilizers include both the $X$ stabilizers of that patch and the $X$ stabilizers surrounding the red stabilizer.
    If any of the purple or gray stabilizer checks yields a nontrivial outcome, the state is discarded.
  }
  \label{fig:expansion}
\end{figure*}
For subsequent use in quantum algorithms, it is desirable for the target surface-code qubits to have a large code distance.
However, the logical operators of the $\llbracket 8,3,2 \rrbracket$ code have weight 2, and it is difficult to perform a direct logical $ZZ$ measurement with a large-distance surface code under geometric locality constraints.
To address this issue, we introduce adaptively initialized teleportation (AIT), a tailored initialization procedure for the surface code. Since AIT allows the code distance of the target surface code to be chosen adaptively during teleportation, it enables efficient teleportation from the $\llbracket 8,3,2 \rrbracket$ code to a higher-distance surface code.
The AIT procedure and the ordering of stabilizer measurements are shown in Fig.~\ref{fig:expansion}.

Fig.~\ref{fig:expansion}(a) illustrates the arrangement of product states used to initialize the surface code in the logical $\ket{+}$ state.
A subset of the qubits is initialized in $\ket{0}$, while the remaining qubits are initialized in $\ket{+}$.
In Fig.~\ref{fig:expansion}(b), one $X$ stabilizer is intentionally left unmeasured, yielding an intermediate state suitable for logical $ZZ$ measurements.
In this intermediate state, the logical $Z$ operator is split into two parts: a segment on qubits prepared in $\ket{0}$, whose outcomes are known, and a segment on the $d_Z=3$ patch.
Consequently, only the three qubits prepared in $\ket{+}$ contribute unknown support to the logical $Z$ operator.
Given that the logical $Z$ operator of the $\llbracket 8,3,2 \rrbracket$ code has weight 2, the combined five-qubit support enables a logical $ZZ$ measurement between the two codes.
After performing the logical $ZZ$ measurement, the remaining stabilizers are measured, as shown in Fig.~\ref{fig:expansion}(c), completing the surface code.
If a nontrivial syndrome is measured either in the $d=3$ surface-code region or on an $X$ stabilizer adjacent to the previously unmeasured $X$ stabilizer, the state is discarded. 
In addition, syndrome extraction is repeated three times, and the state is also discarded if a nontrivial syndrome appears in the same region or on the newly measured $X$ stabilizer.
Since the $\llbracket 8,3,2 \rrbracket$ code has distance 2, logical errors arising from three or more faults occur only at higher order and therefore do not affect the leading-order error probability.
By completing the measurement of the missing stabilizers, the surface code attains its full code distance, allowing the logical state to be teleported from the $\llbracket 8,3,2 \rrbracket$ code into a surface code with significantly larger effective distance.

\subsection{Superdense syndrome measurement circuit}
\begin{figure}
  \centering
  \includegraphics[keepaspectratio, scale=0.5]{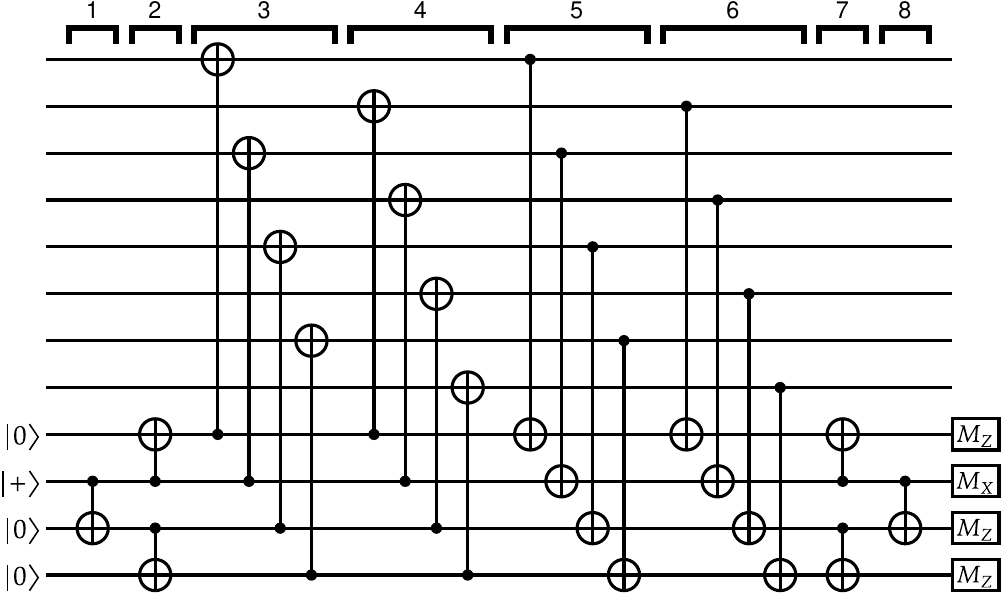}
  \caption{
  \justifying
  Superdense syndrome extraction circuit for the $\llbracket 8,3,2 \rrbracket$ code.
  Using a GHZ state, one $X$ stabilizer and three $Z$ stabilizers are measured simultaneously with a circuit depth of 8.}
  \label{fig:high_density}
\end{figure}

Prior to directly measuring the $\llbracket 8,3,2 \rrbracket$ code, we perform syndrome extraction for both the $X$- and $Z$-type stabilizers.
If any nontrivial syndrome is detected, the state is discarded to preserve fault tolerance.
The syndrome measurements must be executable using only nearest-neighbor interactions on a two-dimensional square lattice, while also keeping the circuit depth as shallow as possible.
To this end, we introduce a superdense syndrome extraction method optimized for the $\llbracket 8,3,2 \rrbracket$ code.
Our method adapts ideas from the parallel syndrome extraction technique of Ref.\cite{lao2020fault} and the high-density measurement scheme developed for color codes in Ref.\cite{gidney2023new}.
The resulting circuit is shown in Fig.~\ref{fig:high_density}.
In this circuit, a GHZ state is prepared and four transversal CNOT layers are applied, enabling the simultaneous measurement of one $X$-type stabilizer and three $Z$-type stabilizers.
The measurement procedure follows the principle of the Hadamard test.
The $X$ stabilizer is inferred from the $X$-basis measurement of the GHZ ancilla, whereas the $Z$ stabilizers correspond to the $Z$-basis outcomes of the individual qubits comprising the GHZ state.
Because three $Z$ stabilizers are measured simultaneously, CNOT gates acting on the data qubits shared among these stabilizers can be consolidated, thereby reducing the circuit depth.
Moreover, each stabilizer measurement inherently serves as both a syndrome extraction and a hook-error detection step.
Since nontrivial syndromes lead to discarding the state, it is unnecessary to distinguish between data errors and hook errors.
With these optimizations, the syndromes of the $\llbracket 8,3,2 \rrbracket$ code can be extracted fault-tolerantly with a circuit depth of only 8.

\section{NUMERICAL SIMULATION}
\label{sec:numerical_simulation}
The circuit proposed in this work uses eight non-Clifford $T/T^\dagger$ gates.
However, full state-vector simulation becomes computationally impractical as the number of qubits grows, due to the exponential scaling of memory and runtime.
To enable efficient numerical evaluation, we replace all $T/T^\dagger$ gates with the identity, thereby converting the entire circuit into a Clifford circuit suitable for simulation using the stabilizer simulator Stim~\cite{gidney2021stim}.
This approximation is appropriate for the performance evaluation in this study.
Under conjugation by $T$, Pauli errors transform as $TZT^\dagger = Z, TXT^\dagger = (X+Y)/\sqrt{2}, TYT^\dagger = (Y-X)/\sqrt{2}$, so the $T/T^\dagger$ layer locally mixes $X$ and $Y$-type errors but does not convert them into $Z$-type errors.
Moreover, since the $T/T^\dagger$ layer consists only of single-qubit gates, it does not enlarge the support of any fault.
These $X,Y$-type components are subsequently detected and removed by the weight-4 $X$-type checks of the $\llbracket 8,3,2 \rrbracket$ code, and therefore do not contribute to the leading-order logical failure events in accepted runs.
To verify this approximation, we used Qulacs~\cite{suzuki2021qulacs} to numerically simulate the full distillation circuit both with and without the transversal $T/T^\dagger$ layer, corresponding to the preparation of $\overline{CCZ}\ket{+++}$ and $\ket{+++}$, respectively. 
We compared the logical error rates obtained from the encoding and syndrome measurement stages of the $\llbracket 8,3,2 \rrbracket$ code in both cases and found no appreciable difference, confirming the validity of the Clifford-circuit approximation used in our simulations.
On the other hand, this replacement modifies the teleported state: instead of teleporting logical $\overline{CCZ}\ket{+++}$, the effective circuit teleports logical $\ket{+++}$.
Consequently, in the effective model, the logical $ZZ$-measurement outcomes no longer influence the final output state.
Logical error mechanisms associated with these measurement outcomes are therefore absent from the simulation, making the estimated logical error rate slightly optimistic.
However, fault patterns originating from these logical $ZZ$ measurements are expected to constitute only a very small fraction of all malignant fault patterns in the full circuit, since they are confined to the teleportation-measurement region. Therefore, even if included, their contribution would only slightly increase the number of malignant fault patterns and would lead at most to a modest correction to the prefactor of the logical error rate.

For the simulations, we adopt the following noise model:
single-qubit gates and idle operations are followed by single-qubit depolarizing noise with probability $p$;
state initialization and measurement are followed by single-qubit flip noise with probability $p$;
two-qubit gates are followed by two-qubit depolarizing noise with probability $p$.
A logical error is declared if any of the three output surface-code logical qubits exhibits a logical failure.
We sample six values of $p$ in the range from $10^{-4}$ to $10^{-3}$, performing $10^7$--$10^8$ trials for each value.

\begin{figure}[h]
  \centering
  \includegraphics[keepaspectratio, scale=1.1]{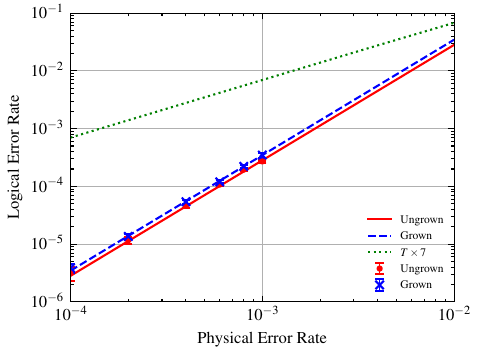}
  \caption{
  \justifying
    Numerical results for the logical error rate $p_L$ as a function of the physical error rate $p$.
    Red corresponds to the case without surface-code expansion, and blue corresponds to the expanded case.
    The solid lines show least-squares fits, yielding $p_L \simeq 300p^2$ in both cases.
    The green dashed line indicates the conventional seven-$T$-gate implementation, for which a logical error occurs if any of the seven $T$ gates fails, giving the leading-order logical error rate $p_L \simeq 7p$.}
  \label{fig:error_rate}
\end{figure}

\begin{figure}[h]
  \centering
  \includegraphics[keepaspectratio, scale=1.1]{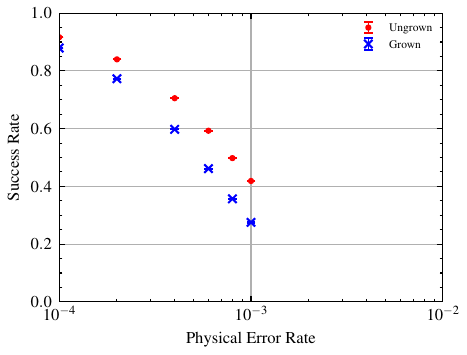}
  \caption{
  \justifying
    Success probability as a function of $p$.
    Red and blue denote the non-expanded and expanded cases, respectively.
    The success probability reaches approximately 30--40\% at $p = 10^{-3}$ and about 90\% at $p = 10^{-4}$.
  }
  \label{fig:success_rate}
\end{figure}

Fig.~\ref{fig:error_rate} displays the logical error rate $p_L$ as a function of the physical error rate $p$: red corresponds to the case without surface-code expansion, and blue corresponds to the case expanded to $d=7$.
Least-squares fitting (solid lines) shows that both cases satisfy $p_L \simeq 300 p^2$.
Compared with the conventional approach, in which a logical error occurs if any of the seven $T$ gates fails, giving a leading-order logical error rate of $p_L \simeq 7p$, the proposed method achieves a substantially lower logical error rate.
At $p = 10^{-3}$, the logical error rate is approximately $p_L \simeq 3\times 10^{-4}$, representing a one-order-of-magnitude reduction;
at $p = 10^{-4}$, the reduction reaches two orders of magnitude, yielding $p_L \simeq 10^{-6}$.
Fig.~\ref{fig:success_rate} shows the success probability: the red and blue curves again represent the non-expanded and expanded cases, respectively.
We observe success probabilities of 30--40\% at $p = 10^{-3}$ and around 90\% at $p = 10^{-4}$, indicating that high-fidelity $CCZ$ states can be obtained with high probability.

Next, at a physical error rate of $p=10^{-3}$, we compare the space-time overhead of the proposed method with that of existing approaches.
As baselines, we consider three representative protocols:
(i) constructing a $CCZ$ state from four $T$ states prepared via zero-level distillation,
(ii) implementing the $CCZ$ gate using seven $T$ states~\cite{nielsen2010quantum} obtained via zero-level distillation, and
(iii) directly distilling a $CCZ$ state using eight $T$ gates.

\begin{figure}[h]
  \centering
  \vspace{2mm}
  \raisebox{6mm}{\includegraphics[keepaspectratio, scale=0.8]{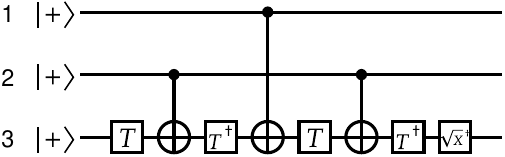}}
  \includegraphics[keepaspectratio, scale=0.5]{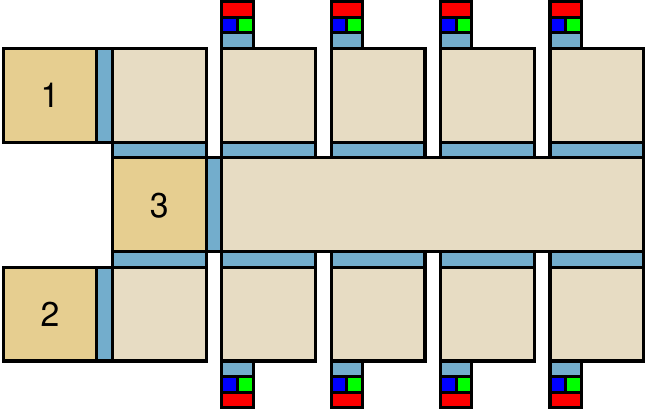}
  \caption{
  \justifying
  Circuit structure and logical-qubit mapping for the method that constructs a $CCZ$ state using four $T$ states produced by zero-level distillation.
  Because four applications of $T/T^\dagger$ are required on the third logical qubit, multiple zero-level distillation factories must be prepared and connected to that qubit.
  The logical-qubit mapping shown here corresponds to an example in which eight zero-level distillation factories are operated in parallel to improve the success probability.}
  \label{fig:4T}
\end{figure}

Zero-level distillation can generate a single $T$ magic state with a spatial overhead of approximately one logical qubit.
The performance data for zero-level distillation are taken from Ref.~\cite{itogawa2025efficient}.
Fig.~\ref{fig:4T} shows the circuit structure and logical-qubit layout for the protocol (i).
In protocol (i), four applications of $T/T^\dagger$ are required on the third logical qubit, and therefore multiple zero-level distillation factories must be run in parallel and connected to that qubit.
For protocol (ii), we assume that the $CCZ$ gate is implemented by sequentially applying seven magic states generated via zero-level distillation.
In this protocol, neither dedicated $CCZ$-magic-state preparation nor gate-teleportation operations for the $CCZ$ gate are required.
For protocol (iii), we use the performance data reported by Gidney et al.~\cite{gidney2019efficient}.

\begin{figure}[h]
  \centering
  \includegraphics[keepaspectratio, scale=1]{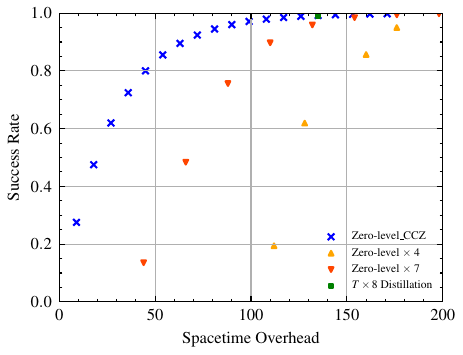}
  \caption{
  \justifying
  Comparison of space-time overhead and success probability for zero-level $CCZ$ state distillation and existing methods.
  The horizontal axis shows the space-time overhead, defined as the product of the number of logical qubits and the number of surface-code syndrome-measurement rounds, and the vertical axis shows the success probability; each point corresponds to a different degree of parallelism.
  The proposed method reduces the required space-time overhead by approximately a factor of 5--10 compared with the existing approaches.
  For the method using eight $T$ states (green point), only a single non-parallelized point is shown, because parallelization makes the space-time overhead excessively large while providing almost no improvement in success probability.
  Moreover, protocols based on generating a $CCZ$ state require gate teleportation, whereas the seven-$T$-gate (red points) implementation directly applies the $CCZ$ gate and therefore does not require teleportation.
  To ensure a fair comparison, we include two additional rounds of syndrome measurement in the space-time overhead for all $CCZ$-state-generation protocols, while no such overhead is added for the seven-$T$-gate approach.
  }  
  \label{fig:simulation_para}
\end{figure}

Fig.~\ref{fig:simulation_para} presents the comparison, with the horizontal axis showing the space-time overhead, defined as the product of the number of logical qubits and the number of surface-code syndrome-measurement rounds and the vertical axis showing the success probability.
We assume an operational regime in which multiple distillation circuits are executed in parallel and only the successful outputs are retained.
In this setting, increasing the degree of parallelism improves the success probability, but also increases the total space-time overhead, generating a clear trade-off.
Each point in the figure corresponds to a different degree of parallelism, illustrating the relationship between the success probability and the resulting overhead.
The space-time overhead is defined as the product of the spatial overhead, given by the number of logical qubits used, and the temporal overhead, given by the number of surface-code syndrome-measurement rounds.
Moreover, generating a $CCZ$ state and directly applying a $CCZ$ gate differ in whether gate teleportation is required.
To ensure fairness, we add two rounds of syndrome measurement to the former to account for the teleportation step.
The results show that the proposed zero-level $CCZ$ state distillation protocol reduces the space-time overhead by approximately a factor of 5--10 compared with existing approaches.
Furthermore, even when the degree of parallelism is increased to improve the success probability, the proposed method consistently achieves a given success probability with a smaller space-time overhead than the competing schemes.

At a physical error rate of $p = 10^{-3}$, the logical error rates of the different approaches can be compared quantitatively.
The proposed zero-level $CCZ$ method scales as $p_L \simeq 300p^2$, yielding $p_L \simeq 3 \times 10^{-4}$.
Since each $T$ state produced by zero-level distillation has a logical error rate of $p_L \simeq 100p^2$, protocols (i) and (ii) accumulate logical errors from four and seven $T$ gates, respectively, yielding $p_L \simeq 4 \times 10^{-4}$ and $p_L \simeq 7 \times 10^{-4}$.
Thus, the proposed method achieves a comparable or lower logical error rate than these approaches.
On the other hand, protocol (iii), the eight-$T$ distillation of Ref.~\cite{gidney2019efficient}, scales as $p_L \simeq 28p^2$, resulting in $p_L \simeq 2.8 \times 10^{-5}$, which is roughly one order of magnitude better than our method.
However, as shown in Fig.~\ref{fig:simulation_para}, this advantage in logical error rate comes at the cost of substantially larger space-time overhead.
Consequently, when both the logical error rate and the space-time overhead are taken into account, the proposed zero-level $CCZ$ distillation offers a more favorable trade-off, making it particularly suitable for early fault-tolerant quantum computation where spatial resources are severely limited.

\section{CONCLUSION}
\label{sec:conclusion}
In this work, we have proposed a zero-level $CCZ$ state distillation circuit that can be implemented using the $\llbracket 8,3,2 \rrbracket$ code under the nearest-neighbor constraints of a two-dimensional square lattice.
Our construction introduces a non-fault-tolerant encoding circuit for the state $\ket{+++}$ and builds upon it to realize a new fault-tolerant encoding procedure.
The non-fault-tolerant encoding circuit and the transversal-CNOT-based encoding method proposed in this work are not limited to the $\llbracket 8,3,2 \rrbracket$ code; they are also applicable to preparing $\ket{0}^{\otimes n}$ and $\ket{+}^{\otimes n}$ states in general CSS codes.
Therefore, this framework is expected to be useful for state-preparation studies beyond magic-state distillation.
We further developed AIT, which enables efficient transfer from the $\llbracket 8,3,2 \rrbracket$ code to a larger-distance surface code without requiring the surface-code expansion procedure used in previous works.
This achieves low-overhead teleportation while maintaining favorable logical error rates.
The proposed zero-level $CCZ$ state distillation circuit operates with very small space-time overhead.
Spatial overhead consists of only 22 physical qubits for distillation and 3 logical qubits for output, which is significantly smaller than in existing protocols.
The time overhead is likewise minimal: the total circuit depth is 24, corresponding to merely three rounds of surface-code syndrome extraction.
As a result, the overall time overhead required for distillation is limited to the $d$ rounds needed for surface-code state preparation.
Overall, the space-time overhead is reduced by a factor of approximately 5--10 relative to conventional approaches.
Numerical simulations confirm that the proposed circuit suppresses the logical error rate to $p_L \simeq 300p^2$, where $p$ is the physical error rate.
This represents an improvement of one order of magnitude at $p=10^{-3}$ and two orders of magnitude at $p=10^{-4}$ over the conventional scheme that uses seven $T$ gates.
The distillation success probability reaches approximately 30\% at $p=10^{-3}$ and 90\% at $p=10^{-4}$, demonstrating that high-fidelity $CCZ$ states can be obtained with high probability.
Because the proposed circuit has low space-time overhead, multiple distillation blocks can be run in parallel, increasing the overall success rate while keeping the overhead manageable.

Given its low resource requirements, the proposed method is a promising candidate for early-FTQC as a practical means of supplying magic states.
In particular, the ability to generate a $CCZ$ state in a single distillation cycle, whereas conventional methods typically require seven rounds of magic-state distillation, offers substantial gains in algorithmic efficiency under tight spatial constraints.
Nevertheless, several challenges remain before the method can be incorporated into full-scale FTQC architectures.
Because the present distillation circuit targets error suppression at the physical level, there is a fundamental limit to the reduction in logical error rates achievable by this approach alone.
Consequently, supplying $CCZ$ states at the FTQC level will likely require combining the proposed method with additional logical-level distillation procedures, which have yet to be fully explored.
Furthermore, our protocol is designed for superconducting qubits arranged on a two-dimensional square lattice.
Other quantum computing platforms—such as trapped ions or neutral atoms—feature different connectivity and gate implementations, and investigating how the proposed method can be adapted or extended to such architectures is an important direction for future work.

\begin{acknowledgments}
  This work is supported by MEXT Quantum Leap Flagship Program (MEXT Q-LEAP) Grant No. JPMXS0120319794, JST COI-NEXT Grant No. JPMJPF2014, JST Moonshot R\&D Grant No. JPMJMS2061, and JST CREST JPMJCR24I3.
\end{acknowledgments}

\appendix
\section{Detailed Circuit Construction of Zero-Level $CCZ$ Distillation}
\label{sec:circuit}
This figure illustrates the physical-qubit layout and gate operations at each step of the zero-level $CCZ$ state distillation protocol.
Solid lines connecting a circle and a plus sign represent CNOT gates, where the circle denotes the control qubit and the plus sign denotes the target qubit.
Yellow faces indicate $X$-type stabilizers, orange faces indicate $Z$-type stabilizers, and purple faces correspond to logical $ZZ$ measurements performed via lattice surgery.
Blue operations act primarily on data qubits, orange operations on ancilla qubits, and green operations on the output surface-code qubits.
The symbol R denotes a reset operation, and M denotes a measurement; the subscripts $Z$ and $X$ specify the measurement basis.
\begin{figure*}[h]
  \centering
  \includegraphics[keepaspectratio, scale=0.35]{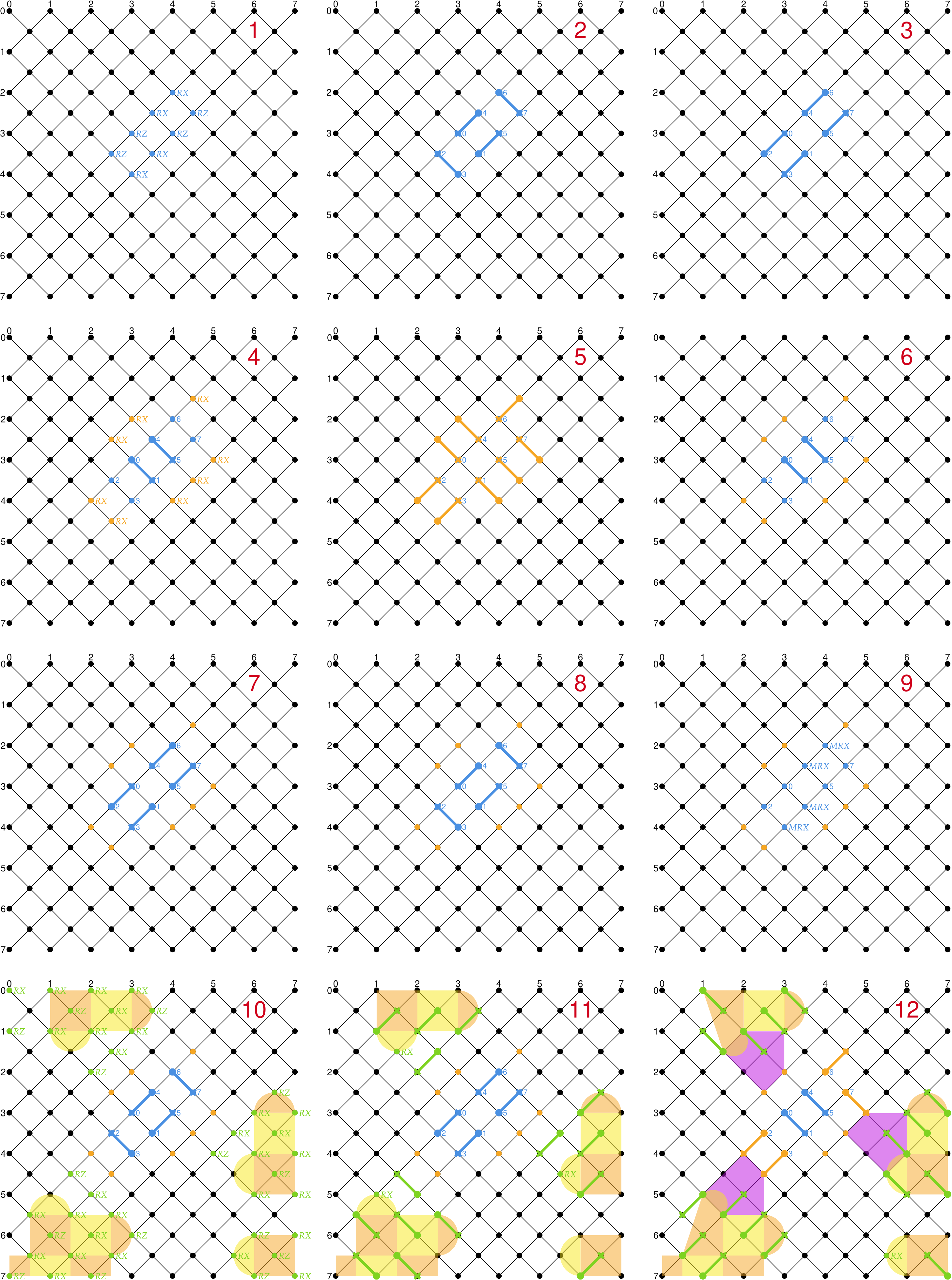}
\end{figure*}
\begin{figure*}[h]
  \centering
  \includegraphics[keepaspectratio, scale=0.35]{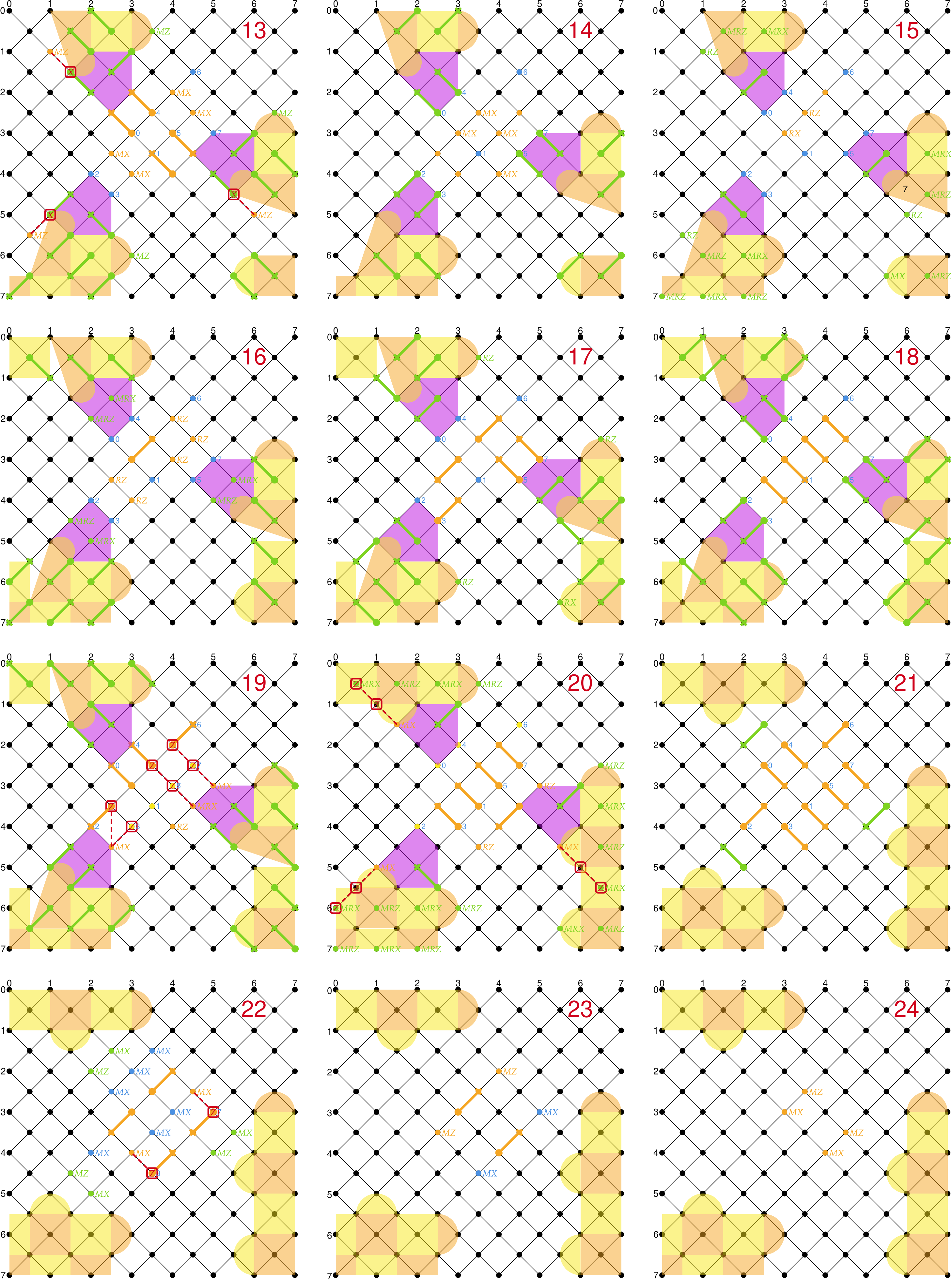}
\end{figure*}

\clearpage
\bibliographystyle{unsrt}
\bibliography{cite}

\end{document}